\documentclass{PoS}

\newcommand{\be}{\begin{equation}}
\newcommand{\ee}{\end{equation}}
\newcommand{\bea}{\begin{eqnarray}}
\newcommand{\eea}{\end{eqnarray}}

\def\slashedD{{\cal D}\hspace{-6pt}\slash}

\title{Aspects of the Bosonic Spectral Action: Successes and Challenges}

\ShortTitle{Aspects of the Bosonic Spectral Action: Successes \& Challenges}

\author{\speaker{Mairi Sakellariadou}\\
        King's College London, University of London, Strand, London WC2R 2LS, U.K.\\
        E-mail: \email{Mairi.Sakellariadou@kcl.ac.uk}}

      \abstract{A short introduction on elements of noncommutative
        geometry, which offers a purely geometric interpretation of
        the Standard Model and implies a higher derivative
        gravitational theory, is presented. Physical consequences of
        almost commutative manifolds are briefly discussed and
        cosmological consequences of the gravitational sector, which
        is shown not to be plagued by linear instability, are
        highlighted. Successes and challenges are discussed. A novel
        spectral action proposal based on zeta function regularisation
        is briefly presented. }

\FullConference{Proceedings of the Corfu Summer Institute 2015 "School and Workshops on Elementary Particle Physics and Gravity"\\

                 1-27 September 2015\\

                 Corfu, Greece}

\begin{document}

\section{Introduction}

Classical cosmology, tested by a variety of precise astrophysical
measurements, is built upon Einstein's theory of General Relativity
and the Cosmological Principle. General Relativity is however a
classical theory, hence its validity breaks down at very high energy
scales. The Cosmological Principle, namely the assumption of a
continuous space-time characterised by homogeneity and isotropy on
large scales, is valid only once we consider late-eras of our
universe, characterised by energies far below the Planck scale. At
very early times, very close to the Big Bang and the so-called Planck
era, quantum corrections can no longer be neglected and geometry may
altogether lose the meaning we are familiar with. To describe the
physics near the Big Bang, a Quantum Gravity theory and the associated
appropriate space-time geometry is hence required.

The available Quantum Gravity proposals can be divided into two
classes. On the one hand, there is String Theory/M-theory, according
which matter consists of one-dimensional objects, strings, which can
be either closed or open (without ends). Different string vibrations
would represent different particles; splitting and joining of strings
would then correspond to different particle interactions. On the other
hand, there are non-perturbative approaches to quantum gravity; some
examples are Loop Quantum Gravity, a Euclidean approach to quantum
gravity like Causal Dynamical Triangulations, and Group Field
Theory. The latter class of models adopts the hypothesis that space is
not infinitely divisible, instead it has a granular structure, hence
it is made out of quanta of space. In the former class of models,
matter is the important ingredient; in the latter one, matter is, so
far, (rather artificially) added.  These two classes of models can be
considered as following a {\sl top-down approach}, whilst they both
inspire cosmological models leading to several observational
consequences.

In the following, I will consider a {\sl bottom-up approach}, in the
sense that I will focus on a proposal attempting to guess the
small-scale structure of space-time near the Planck era, using our
knowledge of well-tested particle physics at the electroweak
scale. More precisely, I will focus on Noncommutative Spectral
Geometry (NCSG). One may argue that at the Planck energy scale,
quantum gravity implies that space-time is a wildly noncommutative
manifold. However, at an intermediate scale, one may assume that the
algebra of coordinates is only a mildly noncommutative algebra of
matrix valued functions, which if appropriately chosen, may lead to
the Standard Model of particles physics coupled to gravity. It is
important to note that according to the NCSG proposal, to construct a
quantum theory of gravity coupled to matter, the gravity-matter
interaction is the most important ingredient to determine the
dynamics; this consideration is not the case for either of the two classes
of proposals mentioned previously.

I will first briefly introduce elements of NCSG and then discuss some of its phenomenological consequences, focusing on its successes and some open questions~\cite{Sakellariadou:2015uda,Sakellariadou:2012jz,Sakellariadou:2010nr}. 

%%%%%%%%%%%%%%%%%%%%%%%%%%%%%%%%%%
\section{NCSG in a nutshell}
%%%%%%%%
Noncommutative spectral geometry~\cite{ncg-book1,ncg-book2,Walterbook}
postulates that the Standard Model (SM) of particle physics is a
phenomenological model which dictates the space-time geometry in order
to get the SM action. In its simplest approach, one considers that at
each point of a four-dimensional Riemannian manifold there is an
internal zero-dimensionality discrete space. Such a Kaluza-Klein type
mildly noncommutative manifold, given by the product ${\cal M}\times
{\cal F}$ of a compact four-dimensional smooth Riemannian spin
manifold ${\cal M}$ and a discrete noncommutative zero-dimensionality
space ${\cal F}$, is called an {\sl almost commutative} manifold.

The main idea we will follow is to characterise Riemannian manifolds by {\sl
  spectral data}, and then apply the same procedure in the case of  almost commutative
manifolds.  Hence, let us first consider a compact four-dimensional
Riemannian spin manifold ${\cal M}$. The set $C^\infty({\cal M})$ of
smooth infinitely differentiable functions forms an algebra ${\cal
  A}=C^\infty({\cal M})$ under point-like multiplication. Then consider
the Hilbert space ${\cal H}=L^2({\cal M},S)$ of square-integrable
spinors $S$ on the spin manifold ${\cal M}$. Note that the algebra
${\cal A}=C^\infty({\cal M})$ acts on the Hilbert space ${\cal
  H}=L^2({\cal M},S)$ as multiplication operators. Finally, consider the Dirac operator
$\slashedD=-i\gamma^\mu\nabla_\mu^S$, acting as a first order
differential operator on the spinors $S$. The canonical triple
$(C^\infty({\cal M}), L^2({\cal M},S), \slashedD)$ encodes the
space-time structure. In addition, we introduce the $\gamma_5$
operator, which is just a $\mathbb{Z}_2$-grading, with $\gamma_5^2=1,
\gamma_5^\star=\gamma_5$. It plays the r\^ole of a chirality operator,
in the sense that it decomposes ${\cal H}$ into a positive and negative
eigen-space: $L^2({\cal M},S)=L^2({\cal M},S)^+\otimes L^2({\cal
  M},S)^-$.  Let us also introduce an antilinear isomorphism
$J_{\cal M}$, playing the r\^ole of a charge conjugation operator on
spinors, with $J_{\cal M}^2=-1,\ J_{\cal M}\slashedD=\slashedD J_{\cal M}, \
J_{\cal M}\gamma_5=\gamma_5 J_{\cal M}$.

In a similar way, the noncommutative space ${\cal F}$, which encodes the
internal degrees of freedom at each point in space-time ${\cal M}$, can
be described by the real spectral triple $(\mathcal{A_F}, \
\mathcal{H_F},\ \mathcal{D_F})$. It lead to a gauge theory on the spin
manifold ${\cal M}$. Here $(\mathcal{A_F}$ is an involution of
operators on the finite-dimensional Hilbert space ${\cal H_F}$ of
Euclidean fermions. The matrix algebra $\mathcal{A_F}$ contains all
information usually carried by the metric. The axioms of the spectral
triples imply that the Dirac operator of the internal space is the fermionic mass matrix. Hence, $\mathcal{D_F}$ is a
$96\times 96$ matrix in terms of the $3\times 3$ Yukawa mixing
matrices and a real constant necessary to obtain neutrino mass terms.
Consider also a grading $\gamma_{\cal F}$, with $\gamma_{\cal F}= + 1$
for left-handed and $\gamma_{\cal F}= - 1$ for right-handed fermions, and
a conjugation operator $J_{\cal F}$:
\[J_{\cal F}= \left( \begin{array}{cc}
    & 1_{48}  \\
    1_{48} & \end{array} \right)~.\] The almost commutative manifold
${\cal M}\times {\cal F}$ is thus given by the spectral triple
$(\mathcal{A}, \mathcal{H}, \mathcal{D})$, with
\begin{eqnarray}
&\mathcal{A}=C^\infty (\mathcal{M}) \otimes \mathcal{A_F} =
C^\infty ( \mathcal{M , A_F})~,\nonumber\\ &\mathcal{H} =
\mathcal{L}^2(\mathcal{ M }, S) \otimes \mathcal{H_F} =
\mathcal{L}^2(\mathcal{M} , S \otimes \mathcal{H_F ) }~,\nonumber\\ & {\cal D} =
\slashedD \otimes \mathbb{I} + \gamma_5 \otimes {\cal D}_{
\mathcal{F}}~.\nonumber 
\end{eqnarray}
The finite dimensional algebra ${\cal A_F}$, which is the main input,
must be in agreement with noncommuative geometry properties, while it
must be chosen such that it can lead to the SM. The appropriate choice is~\cite{Chamseddine:2007bm}
\[ 
\mathcal{A_F}=M_a(\mathbb{H})\oplus M_k(\mathbb{C})~,
\]
with $M_a(\mathbb{H})$ the algebra of quaternions and
$M_k(\mathbb{C})$ the algebra of complex $k\times k$ matrices with
$k=2a$. The lowest allowed value of $k$  in order to obtain 16
fermions in each of the three generations, where the number of
generations is an (external) physical input, is $k=4$.

It is important to note that the choice of an almost commutative
manifold has deep physical implications. The spectral triple $({\cal A,
  H, D}, J, \gamma)$ defining the almost commutative manifold ${\cal
  M}\times{\cal F}$ can be written as
\begin{equation}
({\cal A, H, D}, J, \gamma)=({\cal A}_1, {\cal H}_1, {\cal D}_1, J_1,
  \gamma_1)\otimes({\cal A}_2, {\cal H}_2, {\cal D}_2, J_2,
  \gamma_2)~, \nonumber\\
\end{equation}
with
\begin{equation}
{\cal A}={\cal A}_1\otimes{\cal A}_2~,~{\cal H}={\cal H}_1\otimes{\cal H}_2
~,~ {\cal D}={\cal D}_1\otimes 1 +\gamma_1\otimes{\cal D}_2~,
\gamma=\gamma_1\otimes\gamma_2~,~J=J_1\otimes J_2~,
\nonumber
\end{equation}
where $J^2=-1, [J,{\cal D}]=0, [J_1,\gamma_1]=0$ and $\{J,\gamma\}=0$.\\  The
algebra doubling is strongly related to dissipation, to the the gauge
structure of the SM, whilst it offers a way to generate the seeds of
quantisation~\cite{Sakellariadou:2011wv}.  Moreover, the doubling of
the algebra offers a natural explanation for neutrino mixing, since by
linking the algebra doubling to the deformed Hopf algebra, one can
build Bogogliubov transformations and argue the emergence of neutrino
mixing~\cite{Gargiulo:2013bla}.

In order to extract physical consequences of the NCSG construction,
one needs to obtain a Lagrangian. To do so we will apply the spectral
action principle, according which the bosonic part of the action is of the
form
\[
{\rm Tr}(f({\cal D}^2_A/\Lambda^2))~,
\]
where ${\cal D}_A$ is the fluctuated Dirac operator, $f$ is a cutoff
function (a positive function that goes to zero for large values of
its argument) and $\Lambda$ a cutoff scale, denoting the energy scale
at which the Lagrangian is valid. Hence, the bosonic part of the
action sums up eigen-values of the fluctuated Dirac operator, smaller
than the cutoff energy scale. Since this action depends on the cutoff
energy scale and (mildly) on the cutoff function, we will call it {\it
  the cutoff bosonic spectral action}. One then evaluates the trace with
heat kernel techniques, and writes the bosonic cutoff spectral action
in terms of Seeley-de Witt coefficients. The asymptotic
expansion of the trace thus reads
\begin{equation}
{\rm Tr}(f({\cal D}^2_A/\Lambda^2))\sim 2f_4\Lambda^4 a_0({\cal
  D}_A^2) + 2f_2\Lambda^2 a_2({\cal D}_A^2) + f(0)a_4({\cal D}_A^2)
+{\cal O}(\Lambda^{-2})~,
\end{equation}
in terms of only three of the momenta of the cutoff function $f$,
given by
\begin{equation}
f_4=\int_0^\infty f(u)u^3du\ ,\ f_2=\int_0^\infty f(u)udu\ , f_0=f(0)~.
\end{equation}
related respectively to the cosmological constant, the gravitational constant and
the coupling constants at unification.  Performing a
straightforward but long calculation, one finally writes the cutoff
bosonic spectral action, modulo gravitational terms, as
\begin{eqnarray}
S_\Lambda ={-2a f_2\Lambda^2+e f_0\over \pi^2} \int
|\phi|^2\sqrt{g}d^4x +{f_0\over 2\pi^2}\int a|D_\mu\phi|^2\sqrt{g}d^4x
-{f_0\over 12\pi^2}\int aR|\phi|^2\sqrt{g}d^4x \nonumber\\ -{f_0\over
  2\pi^2}\int \left(g_3^2 G_\mu^i G^{\mu i}+g_2^2F_\mu^a F^{\mu\nu
  a}+{5\over 3}g_1^2B_\mu B^\mu\right) \sqrt{g}d^4x +{f_0\over
  2\pi^2}\int b|\phi|^4\sqrt{g}d^4x +{\cal O}(\Lambda^{-2})~,
\end{eqnarray}
with $a, b, c, d, e$ constants depending on the Yukawa parameters.
Adding to the above bosonic action $S_\Lambda$, the fermionic part
\begin{equation}
(1/2)\langle J\Psi, {\cal D}_A \Psi\rangle~;~ \Psi\in{\cal H}^+~,
\label{fermionic}
\end{equation}
one obtains the full SM Lagrangian. The cutoff scale is set at the Grand Unified Theories (GUT) scale,
since among the relations between the coefficients in the spectral action
one obtains the relation $g_2^2=g_3^2=(5/3)g_1^2$ for the three couplings, valid in the context of several GUTs groups.  Following
a renormalisation group analysis~\cite{ccm} one then obtains
predictions for the SM, which turn out to be in agreement with the most current
experimental data.  In particular, one obtains a Higgs doublet with a
negative mass term and a positive quartic term, which implies that the
electroweak symmetry is spontaneously broken. Note, that current
developments~\cite{Stephan,Chamseddine:2012sw,Devastato:2013oqa,Chamseddine:2013kza}
of the noncommutative spectral geometry proposal are in agreement with
the experimentally found Higgs mass.

%%%%%%%%%%%%%%%%%%%%%%%%%%%%%%%%
\section{The gravitational sector}

Noncommutative spectral geometry leads to an {\sl extended}
gravitational theory, in the sense that the gravitational sector
includes additional terms beyond the ones of the Einstein-Hilbert action.  The
gravitational part of the cutoff bosonic spectral action, in Euclidean
signature, reads~\cite{ccm}
\bea
 S_{\rm gr}^{\rm E} &=& \int \left(
\frac{1}{2\kappa_0^2} R + \alpha_0
C_{\mu\nu\rho\sigma}C^{\mu\nu\rho\sigma} + \gamma_0 +\tau_0 R^\star
R^\star
\right.  
%\nonumber\\
+ \frac{1}{4}G^i_{\mu\nu}G^{\mu\nu
  i}+\frac{1}{4}F^\alpha_{\mu\nu}F^{\mu\nu\alpha}
\nonumber\\
&&\ \ \ \ \ \ +\frac{1}{4}B^{\mu\nu}B_{\mu\nu} +\frac{1}{2}|D_\mu{\bf
  H}|^2-\mu_0^2|{\bf H}|^2
%\nonumber\\ 
\left.
- \xi_0 R|{\bf H}|^2 +\lambda_0|{\bf H}|^4
\right) \sqrt{g} \ d^4 x~, \eea
where 
\bea
\kappa_0^2=\frac{12\pi^2}{96f_2\Lambda^2-f_0\mathfrak{c}}~&,&~
\alpha_0=-\frac{3f_0}{10\pi^2}~,\nonumber\\
\gamma_0=\frac{1}{\pi^2}\left(48f_4\Lambda^4-f_2\Lambda^2\mathfrak{c}
+\frac{f_0}{4}\mathfrak{d}\right)~&,&~
\tau_0=\frac{11f_0}{60\pi^2}~,\nonumber\\
\mu_0^2=2\Lambda^2\frac{f_2}{f_0}-{\frac{\mathfrak{e}}{\mathfrak{a}}}~&,&~
\xi_0=\frac{1}{12}~,\nonumber\\
\lambda_0=\frac{\pi^2\mathfrak{b}}{2f_0\mathfrak{a}^2}~&,&~
{\bf H}=(\sqrt{af_0}/\pi)\phi~, \eea
with ${\bf H}$ a rescaling of the Higgs field $\phi$ to normalize the
kinetic energy, and $\mathfrak{a,b,c,d,e}$ parameters related to the
particle physics model. Let me remark that this action has to be seen {\it \`a la Wilson}.
We have hence obtained the Einstein-Hilbert action with a cosmological
term, a conformal Weyl term and a conformal coupling of the Higgs
field to the background geometry. Note that the fourth term is a
topological term related to the Euler characteristic of the space-time
manifold; thus a nondynamical term. In addition to the gravitational sector written above, one obtains also the SM action.

To use the gravitational part of the bosonic spectral action, one must assume that it is also valid
for a Lorentzian manifold. The next question one has to address is
whether the gravitational sector of the bosonic spectral action,
leading to a fourth-order gravitational theory, is not plagued by
linear instability. Considering the spectral action within a
four-dimensional manifold with torsion, one can show that in the
vacuum case, the equations of motion reduce to the second order
Einstein's equations, implying linear
stability~\cite{Sakellariadou:2016dfl}.  To address the nonvacuum case,
one can then consider the spectral action of an almost commutative
geometry and show that the Hamiltonian is bounded from below, securing
also in this case, which is of interest to us here, the linear stability of the
theory~\cite{Sakellariadou:2016dfl}.

The equations of motion that one obtains from gravitational part of
the asymptotic expression of the cutoff bosonic spectral action
are~\cite{Nelson:2008uy}
\be\label{eq:EoM2} R^{\mu\nu} - \frac{1}{2}g^{\mu\nu} R +
\frac{1}{\beta^2} \delta_{\rm cc}\left[
  2C^{\mu\lambda\nu\kappa}_{;\lambda ; \kappa} +
  C^{\mu\lambda\nu\kappa}R_{\lambda \kappa}\right]= 
%\ 8\pi G
\kappa_0^2 \delta_{\rm cc}T^{\mu\nu}_{\rm matter}~, \ee
with
\be
\beta^2 \equiv -\frac{1}{4\kappa_0^2 \alpha_0}
\ \ \ \ \mbox{and}\ \ \ \
\delta_{\rm cc}\equiv[1-2\kappa_0^2\xi_0{\bf H}^2]^{-1}~,
\ee
where $\beta^2$ (or equivalently, $\alpha_0$, or $f_0$) is related to the
coupling constants at unification, and $\delta_{\rm cc}$ encodes the
conformal coupling between the Higgs field and the Ricci scalar.  Neglecting the nonminimal coupling (namely, setting
$\delta_{\rm cc}=1$) one can show that since the Weyl tensor vanishes for a
Friedmann-Lema\^{i}tre-Robertson-Walker space-time, then in such a case there are
no corrections to Einstein's equations~\cite{Nelson:2008uy}.  Any
modifications at leading order will only arise for anisotropic and
inhomogeneous geometries~\cite{Nelson:2008uy}.

As energies however increase, one may no longer set $\delta_{\rm cc}=1$,
and the corresponding background equations are~\cite{Nelson:2008uy}
\be R^{\mu\nu} - \frac{1}{2}g^{\mu\nu}R =
\kappa_0^2\left[\frac{1}{1-\kappa_0^2 |{\bf H}|^2/6}\right] T^{\mu\nu}_{\rm
  matter}~, \ee 
where we have set $\beta=0$, just for simplicity.  Thus one observes
that the nonminimal coupling between the Higgs field and the Ricci
scalar, leads to an effective gravitational constant, or equivalently,
or equivalently, to an enhancement of the self-interaction of the
Higgs field.

Given that the model has no freedom to introduce extra scalar fields,
one may wonder whether the Higgs field, through its nonminimal
coupling to the background geometry, could play the r\^ole of the
inflaton\cite{Nelson:2009wr,Buck:2010sv}. To address this question one
looks for a flat region of the Higgs potential.  Considering the
renormalisation of the Higgs self-coupling up to two-loops, one finds
that for each value of the top quark mass, there is a value of the
Higgs mass where the Higgs potential is locally
flattened~\cite{Buck:2010sv}. However, the flat region is very narrow
and to achieve a sufficiently long inflationary era, the slow-roll
must be very slow, leading to an amplitude of density perturbations
incompatible with Cosmic Microwave Background data
(CMB)~\cite{Buck:2010sv}.

Finally, one can study the effects of NCSG in a perturbed
background. Let us consider linear perturbations
$g_{\mu\nu}=\eta_{\mu\nu}+\gamma_{\mu\nu}$ around a Minkowski
background $\eta_{\mu\nu}$. The linearised equation of motion then
reads~\cite{Nelson:2010rt}
\begin{equation}
    \left(1-\frac{1}{\beta^2}\Box_\eta \right)\Box_\eta {\bar
    h}^{\mu\nu}= - 2\kappa^2 T^{\mu\nu}_{\rm matter}\,,
\end{equation}
where $\kappa^2\equiv 8\pi G$ and $\beta^{2}
=\displaystyle{5\pi^2/(6\kappa^2f_0)}$. Note that $T^{\mu\nu}_{\rm
  matter}$ is taken to lowest order in $\gamma^{\mu\nu}$. To write the
above equation, we have defined the tensor
\be
  {\bar h}_{\mu\nu}={\bar \gamma}_{\mu\nu}-\frac{1}{3\beta^2}\,
  Q^{-1}\left(\eta_{\mu\nu}\Box_\eta-\partial_\mu\partial_\nu\right)\gamma\,,
\ee
with
\be Q\equiv 1-\frac{1}{\beta^2}\, \Box_\eta \ \ \ \ \mbox{and}\ \ \ \ {\bar
  \gamma}_{\mu\nu}=\gamma_{\mu\nu}-{1\over 2}\eta_{\mu\nu}\gamma~.
\ee
One can then impose constraints on $\beta$ from astrophysical data.
More precisely, one may restrict $\beta$ (and therefore $f_0$) by
requiring that the energy lost to gravitational radiation by binary
pulsar systems agrees with the (standard) General Relativity prediction within observational
uncertainties. Such consideration implied a (rather weak) limit, namely
$\beta \gtrsim 7.55\times 10^{-13} {\rm m}^{-1}$~\cite{Nelson:2010ru},
which however can be improved if in the future data from rapidly
orbiting nearby binaries become available.

Considering data from the Gravity Probe B and the LARES experiments,
the limit on $\beta$ has been improved, namely $\beta\gtrsim 7.1\times
10^{-5} {\rm m}^{-1}$ and  $\beta\gtrsim
1.2\times 10^{-6} {\rm m}^{-1}$,
respectively~\cite{Lambiase:2013dai,Capozziello:2014mea}.

The tighter constraint on the parameter $\beta$ can be set using
torsion balance experiments. It turns out that the
modifications to the Newton potentials induced by the spectral action
are similar to those due to a fifth-force potential. One thus finds
$\beta \gtrsim 10^4 \mbox{m}^{-1}$, which is by far the strongest
limit~\cite{Lambiase:2013dai}.

%%%%%%%%%%%%%%%%%%%%%%%%%%%%%%%%%%%%%%%%%%
\section{The zeta function regularisation}

The cutoff bosonic spectral action, defined and explored previously,
has certainly several merits. It leads to a description of geometry in
terms of spectral properties of operators and can provide an
explanation of the most successful particle physics we have, namely
the Standard Model.  Since not all gauge groups can fit into the
framework, one may conclude that absence of large groups (like SO(10))
prevents proton decay; hence an encouraging outcome of the proposal.
However, the meaning of the cutoff scale remains unclear, the
dimensional parameters appear with incorrect values (a {\sl hierarchy
  problem}), and there is also a (mild) dependence on the cutoff
function. Moreover, and maybe this is the most important concern, the
asymptotic expansion, valid only in the weak-field approximation,
invalidates the theory in the ultraviolet regime, when one expects
noncommutative geometry to play an important r\^ole.

To address such issues, the {\sl zeta bosonic spectral action}
\be S_\zeta \equiv \lim_{s\rightarrow 0} \rm{Tr} {\cal D}^{-2s}\equiv
\zeta(0,{\cal D}^2)~, \ee
has been proposed~\cite{Kurkov:2014twa}. It is just the $a_4$ heat
kernel coefficient associated with the Laplace type operator ${\cal
  D}^2$:
\be S_\zeta = a_4\left[{\cal D}^2\right] = \int d^4 x \,\sqrt{g}\, L ~~
\mbox{with} ~~ L(x) = a_4({\cal D}^2,x)~,  \ee
and leads to the Lagrangian density
\bea L(x) &=& \alpha_1M^4 + \alpha_2 M^2 R +\alpha_3M^2H^2
+ \alpha_4 B_{\mu\nu}B^{\mu\nu} + \alpha_5
W_{\mu\nu}^{\alpha}W^{\mu\nu\,\alpha} + \alpha_6
G_{\mu\nu}^aG^{\mu\nu\,a} \nonumber\\ && +
\alpha_7\,H\left(-\nabla^2-\frac{R}{6}\right)H + \alpha_8 H^4 +
\alpha_9 C_{\mu\nu\rho\sigma}C^{\mu\nu\rho\sigma} +
\alpha_{10}R^*R^*~, \eea
where $B_{\mu\nu}$, $W_{\mu\nu}$ and $G_{\mu\nu}$ are respectively the
field strength  tensors of the U(1), SU(2) and SU(3) gauge fields; the
coefficients $\alpha_{1},..,\alpha_{10}$ are dimensionless constants
determined by the Dirac operator, the term $R^*R^*$ is the
Gauss-Bonnet density, and $C$ denotes the Weyl tensor. Note that the 
$\alpha_{1}, \alpha_2, \alpha_3$ coefficients cannot be taken by the spectral action, i.e. the lower-dimensional operators must be normalised by hand.

It is worth noting that the dimensionful quantity $M$ corresponding to
the Majorana mass of the right-handed neutrino in the Dirac operator,
is considered here as being constant. Since for $M = 0$ there are no
dimensionful constants in the bare Lagrangian, one concludes that the
cosmological constant, the Higgs mass parameter and the gravitational
constant would not arise from renormalisation.  A nonzero element in
the Dirac operator corresponding to a neutrino Majorana mass is also
essential in order to get the experimentally found Higgs mass. Such a
term in the Dirac operator can be either $a_i\psi^c\sigma(x)\psi$
(with $i$ a generation index, $i=1,2,3$), or of the more general form
$\psi^c(a_i\sigma(x)+M_1)\psi$ with $a_i, M_i$ constants for right-
and left-handed neutrinos in the three generations. In the former
case, no dimension zero and two operators appear in the classical
action; thus one will have to achieve dynamical generation of the
three scales upon quantisation. In the latter case, one chose to use
constant terms in order to introduce the $M^4, M^2 H^2, M^2R$ terms in
the action, and consequently get the respective counter terms via
ultraviolet renormalisation.

Following the zeta spectral action, one has no higher (than 4)
dimensional operators, hence the theory is renormalisable and
local. In addition, there are no issues with asymptotic expansion and
convergence. The zeta bosonic spectral action is purely spectral with
no dependence on a cutoff function.  Moreover, while in the context of
the zeta spectral action, the gravitational spectral dimension is
equal to 2, implying that the gravitational propagation decreases
faster at infinity due to the presence of fourth-derivatives, within
the cutoff spectral action, the spectral dimension vanishes for all
spins~\cite{Alkofer:2014raa}.

%%%%%%%%%%%%%%%%%%%%%%%
\section{Conclusions}
Noncommutative spectral geometry on the one hand addresses conceptual
issues of the SM, while on the other hand it offers a geometrical
framework to study physics at the quantum gravity regime.  In the
noncommutative spectral geometry framework, gravity and the Standard
Model fields are put together into geometry and matter on a
Kaluza-Klein noncommutative space.  Then making use of known
experimental data at the well-tested electroweak scale, one tries to
understand the small-scale space-time structure. Applying the spectral
action within an almost commutative manifold, one gets gravity
combined with Yang-Mills and Higgs. Hence, this approach offers a
purely geometric interpretation of the SM coupled to gravity, and in
addition it offers a natural framework to address early universe cosmology.

To address physical consequences of the spectral action one admits its
validity in the Lorentzian signature; an important issue that deserves
further investigation. Within the context of the cutoff bosonic
spectral action, an important remaining open issue is the weak-field
approximation, in the sense that the expansion in reverse orders of
the cutoff scale is only valid when fields and their derivatives are
smaller than the cutoff scale. In the context of the zeta spectral
action, one may have to find a dynamical generation of the three
dimensionful fundamental constants, namely the cosmological constant,
the Higgs vacuum expectation value and the gravitational constant.

It remains an open question of whether inflation, if at all needed
within a wildly nocnommutative manifold, can be naturally incorporated.
The known scalar fields, appearing in the NCSG action, could provide
through their nonminimal coupling to the background geometry an era of
accelerated expansion but fail to match the cosmic microwave
background temperature anisotropies data. Unfortunately, the
successful $R^2$-type inflation~\cite{Starobinsky}, favoured by the
Planck CMB~\cite{Ade:2015lrj} data, cannot be applied in the higher
derivative gravitational theory obtained by noncommutative spectral
geometry. It is not clear yet whether one can accommodate a
dilaton-type inflation~\cite{Chamseddine:2005zk} or use a scalar
field, in a beyond the Standard Model scenario like the Pati-Salam
model~\cite{Chamseddine:2015ata}, as a successful inflaton candidate.

%%%%%%%%%%%%%%%%%%%%%%%%%%

\end{document}